\documentclass[10pt,twocolumn,oneside,journal]{IEEEtran}
\usepackage{cite}
\usepackage{graphicx}
\usepackage{amsmath}
\usepackage{times}
\usepackage{latexsym}
\usepackage{graphicx}
\usepackage{epstopdf}
\usepackage{bm}
\usepackage{amssymb}
\usepackage[center]{caption2}
\usepackage{stfloats}
\usepackage{cases}
\usepackage{array}
\usepackage{setspace}
\usepackage{fancyhdr}
\usepackage{fancyhdr}
\usepackage{algorithm}
\usepackage{algpseudocode}
\usepackage{textcomp}
\usepackage{wasysym}
\usepackage{url}
\usepackage{xcolor}
\usepackage{verbatim}
\usepackage{stfloats}
\usepackage{float}

\allowdisplaybreaks[4]

\newcaptionstyle{mystyle1}{%
  \centering TABLE  \captiontext \par}
\captionstyle{mystyle1}
\newcaptionstyle{mystyle2}{%
  \captionlabel $.$ \: \captiontext \par}
\captionstyle{mystyle2}
\newcaptionstyle{mystyle3}{%
  \captionlabel $.$ \: \captiontext \par}
\captionstyle{mystyle3}

\begin{document}

\title{Graph-based Cooperative Caching in Fog-RAN}



\author{Xiaoting~Cui,
Yanxiang~Jiang,~\IEEEmembership{Member,~IEEE},
Xuan~Chen, 
Fu-Chun~Zheng,~\IEEEmembership{Senior~Member,~IEEE},
and Xiaohu You ~\IEEEmembership{Fellow,~IEEE}
\thanks{This work has been presented at IEEE ICNC 2018 Workshops.}
\thanks{X. Cui, Y. Jiang, X. Chen, and X. You are with the National Mobile Communications Research Laboratory, Southeast University, Nanjing 210096, China. (e-mail: yxjiang@seu.edu.cn)}
\thanks{F. Zheng is with the National Mobile Communications Research Laboratory, Southeast University, Nanjing 210096, China, and the Department of Electronic Engineering, University of York, York YO10 5DD, U.K. (e-mail: fzheng@ieee.org)}
}

\maketitle

\begin{abstract}
In this paper, the cooperative caching problem in fog radio access networks (F-RAN) is investigated.
To maximize the incremental offloaded traffic, we formulate the clustering optimization problem with the consideration of cooperative caching and local content popularity, which falls into the scope of combinatorial programming. 
We then propose an effective graph-based approach to solve this challenging problem.
Firstly,  a node graph is constructed with its vertex set representing the considered fog access points (F-APs)  and its edge set reflecting the potential cooperations among the  F-APs. 
Then, by exploiting the adjacency table of each vertex of the node graph,
we propose to get the complete subgraphs through indirect searching for the maximal complete subgraphs  for the sake of a reduced searching complexity.
Furthermore, by using the complete subgraphs so  obtained, a weighted graph
is constructed.
By setting the weights of the vertices of the weighted graph to be the incremental offloaded traffics of their corresponding complete subgraphs,
the original clustering optimization problem can be transformed into
an equivalent 0-1 integer programming problem.
The max-weight independent subset of the vertex set of the weighted graph,
which is equivalent to the objective cluster sets, can then be readily obtained by solving the above optimization problem through the greedy algorithm that we propose.
Our proposed graph-based approach has an apparently low complexity in comparison with the brute force approach which has an exponential complexity.
Simulation results show the remarkable improvements in terms of offloading gain by using our proposed approach.


\end{abstract}

\begin{keywords}
    F-RAN, cooperative caching, clustering, offloaded traffic.
\end{keywords}

\section{Introduction}

With the continuous and rapid proliferation of various intelligent devices and advanced mobile application services, wireless networks have been suffering an unprecedented data traffic pressure in recent years.
Ever-increasing mobile data traffic brings tremendous pressure on capacity-limited backhaul links, especially at peak traffic moments.
Fog radio access networks (F-RAN) as a promising architecture can effectively alleviate the traffic congestion in backhaul links by placing popular contents at fog access points (F-APs) which are equipped with limited caching resources \cite{Peng}.
Due to storage constraint and fluctuant spatiotemporal traffic demands, cooperative caching is an effective way to further increase the offloading gain.


Recently, there have been a lot of  works on cooperative caching.
In \cite{MTao}, a cooperative multicast method  was proposed to minimize {the content-delivery rate via backhaul link,}
which investigated local independent cache placement without taking clustering into consideration.
A cache placement method was presented to {minimize the average delay per request} for multi-cell cooperative networks in \cite{Sun},
which assumed that  clustering of base stations was accomplished in advance.
By optimizing relay clustering  and content caching in a joint manner,
a cooperative caching strategy was developed to minimize the outage probability  in \cite{Zheng},
{which used identical content caching among the relays in each cluster for simplicity.}
An energy-aware cooperative transmission strategy by clustering small cells with independent caching ability was proposed in \cite{Chiang},
where each {small base station} was not allowed to share caches with the others.
{Based on the similarities among users requesting similar contents,}
a user clustering algorithm {which could improve the hit rate} was proposed in \cite{ElBamby}.
Besides, most of the prior works on cooperative caching tend to {exploit global content popularity rather than local content popularity,
{which might not even replicate global content popularity.
Local content popularity indeed reflects user interest at each F-AP and might be different from each other \cite{Atan}.}





Motivated by the aforementioned discussions, we propose a graph-based cooperative caching approach by exploiting local content popularity in F-RAN.
We formulate the clustering optimization problem with the consideration of cooperative caching to maximize the incremental offloading traffic.
To solve the above problem, a node graph and then a weighted graph relevant to the considered F-APs are constructed in sequence.
The objective cluster sets can be obtained by searching for the max-weight independent subset of the vertex set of the weighted graph.
The key in our proposed approach is the equivalent transformation from the original optimization problem to the 0-1 integer programming problem by using graph theory.
Great computational complexity reduction can be achieved by using our proposed graph-based approach.

The rest of this paper is organized as follows.
In Section II, the system model and problem formulation are briefly described.
The proposed graph-based cooperative caching approach is presented in Section III.
Simulation results are shown in Section IV.
Final conclusions are drawn in Section V.

\section{System Model and Problem Formulation}

\begin{figure}[htbp]
\centering 
\includegraphics[width=0.40\textwidth]{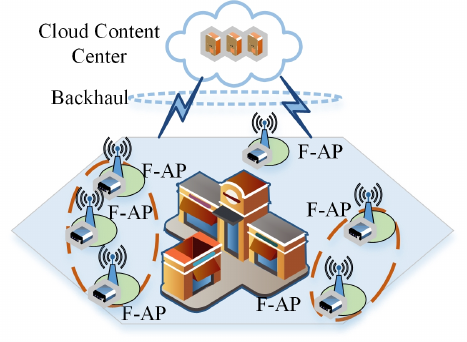}
\caption{Illustration of a cooperative caching scenario.}
\label{model}
\end{figure}

We consider a cooperative caching scenario in a specific region served by ${M}$ F-APs (as shown in Fig. \ref{model}), which  constitute one F-AP set ${\cal M} = \left\{ {1,2,\cdots,m,\cdots,M} \right\}$.
Assume that neighboring F-APs can cooperate with each other and form a cluster to increase the content diversity.
Whether an F-AP can be clustered or not depends on
how well it satisfies some certain rules. 
Assume that the considered $M$ F-APs can constitute one nonclustered set denoted as ${{\cal M}^\text{n}}$ and $N$ disjoint clustered sets denoted as ${{\cal M}^\text{c}_n}$ for $n=1,2, \cdots, N$, and that the set size of ${{\cal M}^\text{c}_n}$ is $S_n$.
Correspondingly, the following relationship can be readily established:
${\cal M} =  \left( { \cup _{n=1}^N}{{{\cal M}^\text{c}_n}} \right) \cup {\cal M}^\text{n} $.
Without loss of generality, we assume that all the  files have the same size of $L$ bits, each F-AP has the same storage space and can store up to $K $ files from the content library ${\cal F} = \left\{ {1,2,\cdots,f,\cdots,F} \right\}$  located in the cloud content center.
{It is assumed that the users in cluster $n$ can be served by the $S_n$ F-APs  to fetch the files of interest from their local caches,
and cluster ${n}$ can cache ${K_n}$ different files.}
Then, we have ${K \le {K_n} \le \min \{ {S_n}K, F\} }$.
Remark here that the local caches of all the F-APs in one cluster will store the same $K$ files if $K_n = K$, and that they will store different $K$ files if $K_n = S_n  K$.
There exists a trade off between files that are popular for the own users of the current F-AP
and files that are frequently requested by the users of the other F-APs in the same cluster.

Let ${p_{m{f}}}$ denote the request probability of file ${f}$ at F-AP $m$ (referred to as local content popularity),
${p_{n{f}}}$  the request probability of file ${f}$ in cluster ${n}$,
and  ${p_{{f}}}$ the  request probability  of file $f $
{in the considered region (referred to as global content popularity).}
Assume that both {local content popularity} and global content popularity
are stationary during the given time period.
Let ${\lambda _m}$ denote the aggregate request arrival rate at F-AP $m$,
and ${w_m} = \lambda _m / {\sum\nolimits_{m = 1}^M {{\lambda _m}} }$
the ratio of the traffic load at F-AP $m$ to the sum load of the $M$ F-APs.
Then, according to \cite{Li}, we have
\begin{equation}
{p_{n{f}}} = \sum\limits_{m \in {{\cal M}^\text{c}_n}} {{p_{m{f}}}\frac{w_m}{\sum\limits_{m \in {{\cal M}_n^\text{c}}} {{w _m}} }}, \
{p_{{f}}} = \sum\limits_{m \in \cal M} {{p_{m{f}}}{w_m}}.
\end{equation}
Without loss of generality, we sort ${p_{m{f}}}$ and ${p_{n{f}}}$  in descending order, and rewrite them as $p_{{mf}}^\text{o}$ and $p_{{nf}}^\text{o}$, respectively.
Then, each F-AP can store popular files to its cache with the request probability in range from $p_{{m1}}^\text{o}$ to $p_{{mK}}^\text{o}$ (for a nonclustered F-AP) or from $p_{{n1}}^\text{o}$ to $p_{{{nK_n}}}^\text{o}$ (for a clustered F-AP).

	Once the requested file can be fetched by  caches of F-APs, the traffic in the backhaul  links can be offloaded.
Let ${T_m}$ denote the average offloaded traffic at single F-AP $m $,  ${T_n^\text{c}}$ the average offloaded traffic in cluster ${n} $, and $T$ the whole offloaded traffic for all the considered $M$ F-APs.
Then, they can be expressed as follows,
\begin{align}
{T_m} &= \sum\limits_{f = 1}^K {{\lambda _m}p_{{mf}}^\text{o} L}, \ 
{T_n^\text{c}} = \sum\limits_{f = 1}^{{K_n}} {\left( {\sum\limits_{m \in {{\cal M}_n^\text{c}}} {{\lambda _m}} } \right)p_{{nf}}^\text{o} L}, \label{S-3} \\
{T} &= \sum\limits_{{n=1} }^N {{T_n^\text{c}}}  + \sum\limits_{m \in {{\cal M}^\text{n}}} {{T_m}}.
\end{align}
Let $T_n^\text{i}$ denote the incremental offloaded traffic of cluster $n$,
and $T^\text{i}$ the  incremental offloaded traffic of all the considered $M$ F-APs.
Then, they can be expressed as follows,
\begin{align}
T_n^\text{i} &=  {{T_n^\text{c}}}   - {\sum\limits_{m \in {{\cal M}_n^\text{c}}}^{} {{T_m}} }, \\
&= \sum\limits_{m \in {{\cal M}_n^\text{c}}} {{\lambda _m}\left( {\sum\limits_{f = 1}^{{K_n}} {p_{{nf}}^\text{o}}  - \sum\limits_{f = 1}^K {p_{{mf}}^\text{o}} } \right)L} , \\
{T^\text{i}} &= \sum\limits_{{n=1}}^N T_n^\text{i}. \label{S-10}
\end{align}
According to the above expressions, the following relationship can be readily established,
\begin{equation}\label{neweq9}
T = {T^{\text i}} + \sum\limits_{m \in {\cal M}} {{T_m}}.
\end{equation}
Note that the second term in the right hand side of \eqref{neweq9}
is unaffected by the clustering strategy of the considered F-APs,
and that to maximize the whole offloaded traffic $T$ is equivalent to maximize the incremental offloaded traffic $T^{\text i}$.
{Therefore, we use the incremental offloaded traffic to reflect the offloading gain of cooperative caching for the sake of reduced computational complexity.}
It can be readily seen that the offloading gain increases with the content diversity and file request probability of the clusters.
Furthermore, the content diversity increases with the F-AP number  in each cluster,
and the request probability varies with the changes of different cooperative F-APs.
Correspondingly, the clustering strategy mainly affects the offloading gain.

To maximize the offloading gain, the clustering strategy should be designed elaborately.
{Firstly, the clusters should be formed by choosing F-APs with closer distance and {greater load difference}.}
The F-APs so selected impose higher interference on their associated users, and are more likely to cooperative with each other \cite{Ashraf}.
Secondly, one F-AP can only be a constituent part of one cluster,
{which ensures exclusive and sufficient usage of its storage space to its serving cluster.}
Based on the above design rules, the clustering optimization problem  can be formulated as follows,
\begin{align}\label{P-1}
&{\mathop {\text{max} }\limits_{\left\{{\cal M}_n^{\text{c}}\right\}_{n=1}^N,{{\cal M}^{\text{n}}}}}  { {T^{\text{i}} } } \\
{\text{s.t.}}\ &{D_{{m}{m '}} \le {\gamma ^\text{d}},\forall {m} \ne {m'} \in {{\cal M}_n^\text{c}},} \tag{\ref{P-1}a}\nonumber\\
&{L_{{m}{m'}} \ge {\gamma ^\text{l}},\forall {m} \ne {m'} \in {{\cal M}_n ^\text{c}},} \tag{\ref{P-1}b} \nonumber\\
&{{{\cal M}_n^\text{c}} \cap {{\cal M}_{n'}^\text{c}} = \varnothing ,\forall {{\cal M}_n^\text{c}} \ne {{\cal M}_{n'}^\text{c}} \subseteq  {\cal M},}  \tag{\ref{P-1}c} \nonumber
\end{align}
where $D_{{m}{m'}} = \left\| {{d_{{m}}} - {d_{{m'}}}} \right\|_2$ denotes the distance between F-AP $m$ and F-AP $m'$ with $d_{{m}}$
representing the geographical coordinate of F-AP $m$ in the Euclidean space,
$L_{{m}{m'}} = \left\| {{\lambda _{{m}}} - {\lambda _{{m'}}}} \right\|_2$ the load difference between F-AP $m$ and F-AP $m'$,
${{\gamma ^\text{d}}}$  the distance threshold,
and ${{\gamma ^\text{l}}}$ the load threshold.

The objective of this paper is to find the optimal clustering strategy by maximizing the offloading gain  with the consideration of  cooperative caching
in F-RAN.


\section{The Proposed Graph-based Cooperative Caching Approach}

This optimization problem in (\ref{P-1}) falls into the scope of combinatorial programming{\cite{Ashraf}}.
A brute force approach is generally required to obtain a globally optimal solution.
However, such a method has an exponential complexity with respect to (w.r.t.)
the number of F-APs {and the number of disjoint cluster sets}, and it is computationally impracticable even for a small size network. Therefore, we commit to an effective graph-based approach to solve this challenging problem.


\subsection{Description of the Proposed Approach}


In our proposed graph-based cooperative caching approach,
firstly, all the considered $M$ F-APs are checked to determine which two of them satisfy the constraints in (\ref{P-1}a) and (\ref{P-1}b).
It is already known that F-APs with appropriate distance and load difference from each other are more likely to cooperate together \cite{Ashraf}.
Then, according to the checking results, the node graph denoted as ${{\cal G}^{\text n}} = ({\cal M},{\cal E})$ is constructed, whose vertex set denoted as $\cal M$  is the F-AP set and whose edge set denoted as $\cal E$ reflects the distance and load difference among the F-APs.
In ${{\cal G}^{\text n}} $, two vertices are connected through an edge if their representing F-APs can cooperate with each other.
Note that one  subgraph of ${{\cal G}^{\text n}}$,  any vertex of which can connect through an edge with a certain vertex in the same subgraph, represents one cluster which consists of a certain number of  cooperative F-APs,
and that one complete subgraph of ${{\cal G}^{\text n}}$, any two vertices of which can connect through an edge,  essentially represents
one candidate cluster of the optimization problem in \eqref{P-1} whose cluster members can cooperative with each other.
We point out here that there may exist a certain vertex not belonging to any subgraph of ${{\cal G}^{\text n}}$, which means that  its representing F-AP is nonclustered.
Fig. \ref{nodegra} illustrates a node graph including 13 vertices as an example.
According to the above descriptions, seeking candidate clusters is equivalent to searching for  complete subgraphs in ${{\cal G}^{\text n}}$.
The algorithm of searching for  complete subgraphs is presented in detail in Section III-B.

Assume that there are $N'$ complete subgraphs that have been obtained through the above searching algorithm.
Let ${\cal H} = \{ {h_1},{h_2}, \cdots ,{h_n}, \cdots ,{h_{N'}}\}$ denote the complete subgraph set so obtained.
{It is clear that $\left\{ {{\cal M}_n^{\text c}} \right\}_{n = 1}^N \subseteq {\cal H}$}.
Then, a weighted graph denoted as {${{\cal G}^{\text w}} = ({\cal H},{\cal B}, \boldsymbol w)$} can be constructed,
where $\cal H$ denotes the vertex set,
$\cal B$ the edge set, 
and $\boldsymbol w$ the weight vector corresponding to the vertices of ${{\cal G}^{\text w}}$ whose elements are set to be the incremental offloaded traffics of their corresponding complete subgraphs (i.e., {${\left[ \boldsymbol w \right]_n} = \mathop T\nolimits_n^{\text i} $}).
{In ${{\cal G}^{\text w}}$, two vertices are connected through an edge if their representing complete subgraphs have a certain identical vertex.}
It is known from graph theory that an independent or stable set is a set of vertices in a graph, {no two of which are adjacent \cite{Bondy}.}
Then, the independent subset of ${\cal H}$ certainly satisfies the constraint in (\ref{P-1}c).
Correspondingly, the objective cluster sets of the optimization problem in \eqref{P-1} can be readily obtained by searching for the equivalent max-weight independent subset of ${\cal H}$ of  the corresponding weighted graph ${{\cal G}^{\text w}}$.
{And the max-weight independent subset of ${\cal H}$ can be obtained by solving a 0-1 integer programming problem,
which will be  presented in detail in Section III-C.}

Remark here that we map one cluster to one complete subgraph,
which guarantees proper-sized clusters and avoids unnecessary intra-cluster signaling overhead,
instead of one connected subgraph as in \cite{Zhou},
{which tends not to constrain the cluster size.}



\begin{figure}[!t]
\centering 
\includegraphics[width=0.4\textwidth]{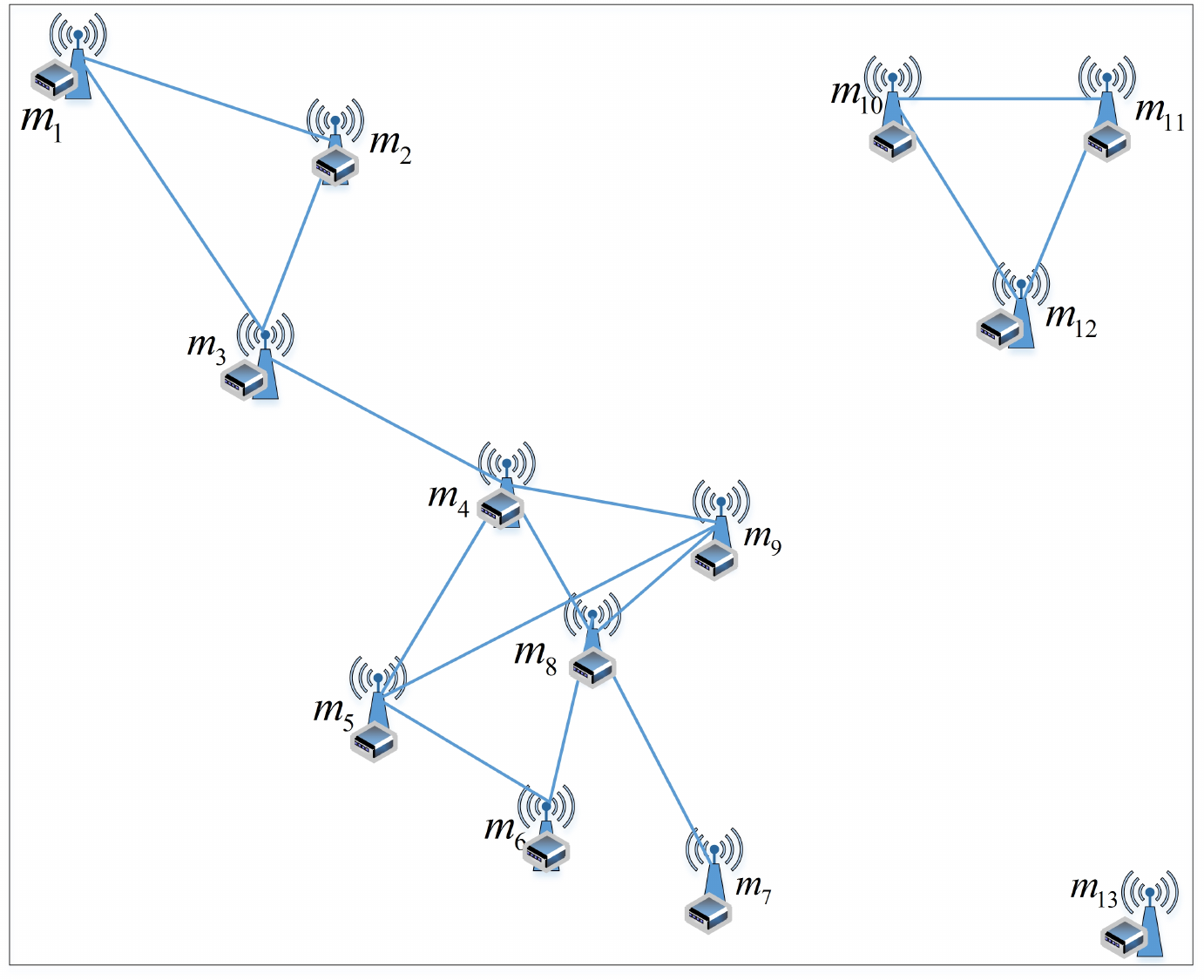}
\caption{Illustration of a node graph including 13 vertices.} 
\label{nodegra}
\end{figure}

\subsection{Searching for Complete Subgraphs}

We propose to search for maximal complete subgraphs to find all the possible complete subgraphs.
It is known from \cite{Bondy} that any complete subgraph must belong to a maximal complete subgraph,
and that it is more difficult  to find complete subgraphs through direct searching than through indirect searching for maximal complete subgraphs. 
We propose to exploit the adjacency table of each vertex in the node graph ${{\cal G}^{\text n}}$ to search for maximal complete subgraphs.
For $m \in {\cal M}$, let
${{\cal T}_m} = \left\{ m \right\} \cup \left\{ {{m'}\left| {m' \in {\cal M},m' > m,\left( {m',m} \right) \in {\cal E}} \right.} \right\}$ denote the adjacency table of vertex $m$ of ${{\cal G}^{\text n}}$,
and ${L_m}$  the table size of ${{\cal T}_m}$.
If {${L_m} = 1$} or ${{\cal T}_m} \subseteq {{\cal T}_{m'}} $ for $m' <m$ and $ m, m' \in {\cal M}$,
it is unnecessary to search for a maximal complete subgraph in ${{\cal T}_m}$.
Remove all the unnecessary or redundant  adjacency tables
{and sort the remaining in descending order denoted as ${\cal T} _{ m }^ {\text o}$ according to their table sizes.}
Let ${{\cal T}}$ denote the set of the reordered adjacency tables.
{Remove any vertex that does not connect with all the other vertices in ${\cal T} _{ m }^ {\text o}$.
Then, the remaining vertices in ${{\cal T}_m ^ {\text o}}$ form a maximal complete subgraph.}
Let ${{\cal G}^{\text m}}$ denote the set of maximal complete subgraphs.
The detailed description of our proposed algorithm of  searching for maximal complete subgraphs is presented in Algorithm \ref{alg-1}.
After maximal complete subgraphs are found, all the possible complete subgraphs can be readily obtained.

\subsection{Searching for Max-Weight Independent Subset}

According to the construction of the weighted graph ${{\cal G}^{\text w}}$,
two vertices in $\cal H$ are adjacent and there exists an edge between them if their representing  candidate cluster sets have some identical elements.
Let ${\boldsymbol x}$ denote the binary indicating vector for the the vertices in $ \cal H$ with ${[\boldsymbol x]_n} = 1$
if the candidate cluster set represented by the vertex $h_n$ belongs to the objective disjoint cluster sets of the original problem in \eqref{P-1}
and ${[\boldsymbol x]_n} = 0$ otherwise.
If the vertices $h_n$ and $h_{n'}$ can be connected through an edge $\left( {{h_n},{h_{n'}}} \right) \in {\cal B}$,
the relationship ${[\boldsymbol x]_n}  [\boldsymbol x]_{n'} = 0$ should be satisfied.
Correspondingly, the original optimization problem in  \eqref{P-1} can be transformed into the following  0-1 integer programming problem,
\begin{align}\label{eq10}
&\mathop {\max }\limits_{\boldsymbol x} { {\boldsymbol w ^T {\boldsymbol x}} }\\
{\text{s.t.}}\ & {{[\boldsymbol x]_n} \in \{ 0,1\}}, \forall h_n \in \cal H, \tag{\ref{eq10}a}\\
&{{{[\boldsymbol x]_n}  {[\boldsymbol x]_{n'}} = 0,\forall \left( {{h_n},{h_{n'}}} \right) \in {\cal B}}}. \tag{\ref{eq10}b}
\end{align}
{The above optimization problem can be solved by linear programming only if its linear relaxation is tight and has a unique integral solution. However, the above two conditions  are hard to be satisfied \cite{Sanghavi}.
Actually, the optimization problem in \eqref{eq10} is a classical problem that maximizes a submodular set function and can often be solved by greedy algorithms \cite{JV}.
Considering that traditional greedy algorithms cannot take full advantage of the specific constraints in (\ref{eq10}a) and (\ref{eq10}b), we then propose a more effective greedy algorithm. }
Let ${{\cal G}^{\text i}}$ denote the independent subset of $\cal H$ and
$w^{\text i}$  the sum weight of all the vertices in ${{\cal G}^{\text i}}$.
Each time we move one vertex with the largest weight from $\cal H$ to ${{\cal G}^{\text i}}$
and remove its adjacent vertices from $\cal H$.
Repeat the above step until $\cal H$ is empty.
The independent subset ${{\cal G}^{\text i}}$ so obtained with the maximum sum weight $w^{\text i}$ is just the max-weight independent subset denoted as $\cal G ^*$ we are searching for.
The detailed description of our proposed greedy algorithm of searching for the max-weight independent subset is presented in Algorithm \ref{alg-2}.

\begin{algorithm}[!t]
	\renewcommand{\algorithmicrequire}{\textbf{Input:}}
	\renewcommand{\algorithmicensure}{\textbf{Output:}}
	\caption{Searching for Maximal Complete Subgraphs}
	\label{alg-1}
	\begin{algorithmic}[1]
		\Require ${{\cal G}^ \text n}$
        \Ensure ${{\cal G}^{\text m}}$            		
        \For {each ${\cal T}_m^{\text o} \in {{\cal T}}$}
        \State Initialize $i=0$,  ${\cal T}_{i}^{\text t}=\varnothing$, ${\cal T'} = {\cal T}_m^{\text o}$;
        \For {each ${j} \in {\cal T'}$}
        \If {${\cal T}_m^{\text o}$ or ${\cal T}_{i'}^ \text t \in \left\{ {{\cal T}_{i'}^\text t} \right\}_{i' = 0}^{i - 1}$ contains both vertex \\
                \qquad \quad $j$ and nonadjacent vertices of $j$}
        \State ${\cal T}_i^\text t = {\cal T}_m^\text o - {{\cal T}_j}$ or ${\cal T}_i^\text t = {\cal T}_{i'}^\text t - {{\cal T}_j}$;
        \State Remove all the nonadjacent vertices of $j$ \\
        \qquad \quad \quad \ \ from the corresponding ${\cal T}_m^{\text o}$ or ${\cal T}_{i'}^\text t$, set \\
        \qquad \quad \quad \ \ $i = i + 1$;
        \EndIf
        \EndFor
        \State ${{\cal G}^{\text m}} = {{\cal G}^{\text m}} \cup {{\cal T}_m^{\text o}} \cup \left\{ {{\cal T}_{i'}^\text t} \right\}_{i' = 0}^{i-1}$
        \EndFor
	\end{algorithmic}
\end{algorithm}

\begin{algorithm}[!t]
	\renewcommand{\algorithmicrequire}{\textbf{Input:}}
	\renewcommand{\algorithmicensure}{\textbf{Output:}}
	\caption{Searching for Max-Weight Independent Subset}
	\label{alg-2}
	\begin{algorithmic}[1]
		\Require ${\cal G} ^ \text w$ 
        \Ensure ${{\cal G}^*}$
        \State Initialize $w^*=0$;	
        \For {each ${h_n} \in {\cal H}$}
        	\State ${{\cal G}^{\text i}} = \left\{ {{h_n}} \right\}$, $w^{\text i} = [\boldsymbol w ]_n$;
        	\State Remove all the adjacent vertices of $h_n$ from $\cal H$;
        	\While {${\cal H} \ne \varnothing $}
                \State Find the vertex with the largest weight from \\
                \quad \quad \quad  $\cal H$  denoted as $h_{{n^{\text max}}}$;
       			\State ${{\cal G}^{\text i}} = {{\cal G}^{\text i}} \cup \left\{ {{h_{{n^{\text max}}}}} \right\}$, $w ^{\text i} = w ^{\text i} + [\boldsymbol w ]_{n^{\text max}}$;
        		\State Remove all the adjacent vertices of $h_{n^{\text max}}$ \\
                  \quad \quad \quad  from $\cal H$;
        	\EndWhile
        	\If {${w^*} < w^{\text i}$}
        		\State ${w^*} = w^{\text i}$, ${{\cal G}^*} = {{\cal G}^{\text i}}$.
        	\EndIf
        \EndFor
	\end{algorithmic}
\end{algorithm}

\subsection{Complexity Analysis}

Let ${\bar L}$ denote the average size of the adjacency tables of all the vertices in the node graph $\cal G^ \text n$.
Then, the computational complexity of searching for maximal complete subgraphs in Algorithm \ref{alg-1} can be calculated to be  ${\mathcal O}(M\bar L)$.
Furthermore, the computational complexity of obtaining all the complete subgraphs can be calculated to be  ${\mathcal O}(P{{\bar S}})$, where  $P$ denotes the number of the maximal complete subgraphs that have been found,
{and ${{\bar S}}$  the average vertex number of all the complete subgraphs.}
Besides, the computational complexity of searching for the max-weight independent subset in Algorithm \ref{alg-2} can be calculated to be ${\mathcal O}(N'{N})$.
In summary, the computational complexity of the proposed approach can be calculated to be  {${\mathcal O}(M\bar L + P{{\bar S}} + N'{N})$.}
By considering  ${\bar L< M}$, ${{\bar S}} < M$ and $N<M$, the computational complexity of the proposed approach is ${\mathcal O}(M^2 + PM + N'M)$ in the worst case.
However, the computational complexity  of the traditional brute force approach can be calculated to be  $\mathcal O ({2^{{N}M}})$.
Obviously, the proposed graph-based approach has a lower complexity in comparison with the brute force approach which has an exponential complexity.

\section{Simulation Results}

In this section, the performance of the proposed graph-based cooperative caching approach is evaluated via simulations.
In the simulations,
we assume the global content popularity $p_f$ follows the Zipf distribution with parameter $z$.
In the following, unless otherwise stated, we set:
$z=0.6$, $M=13$, $F=50$,  
and $L = 200{\rm Mb}$. 
We choose the largest  content diversity (LCD)  \cite{Chen} and uniform-local (U-L) \cite{Liu} approaches as the two  baseline schemes
{which used global content popularity and local content popularity, respectively}.

%
%
%

\begin{figure}[!t]
\centering 
\includegraphics[width=0.48\textwidth]{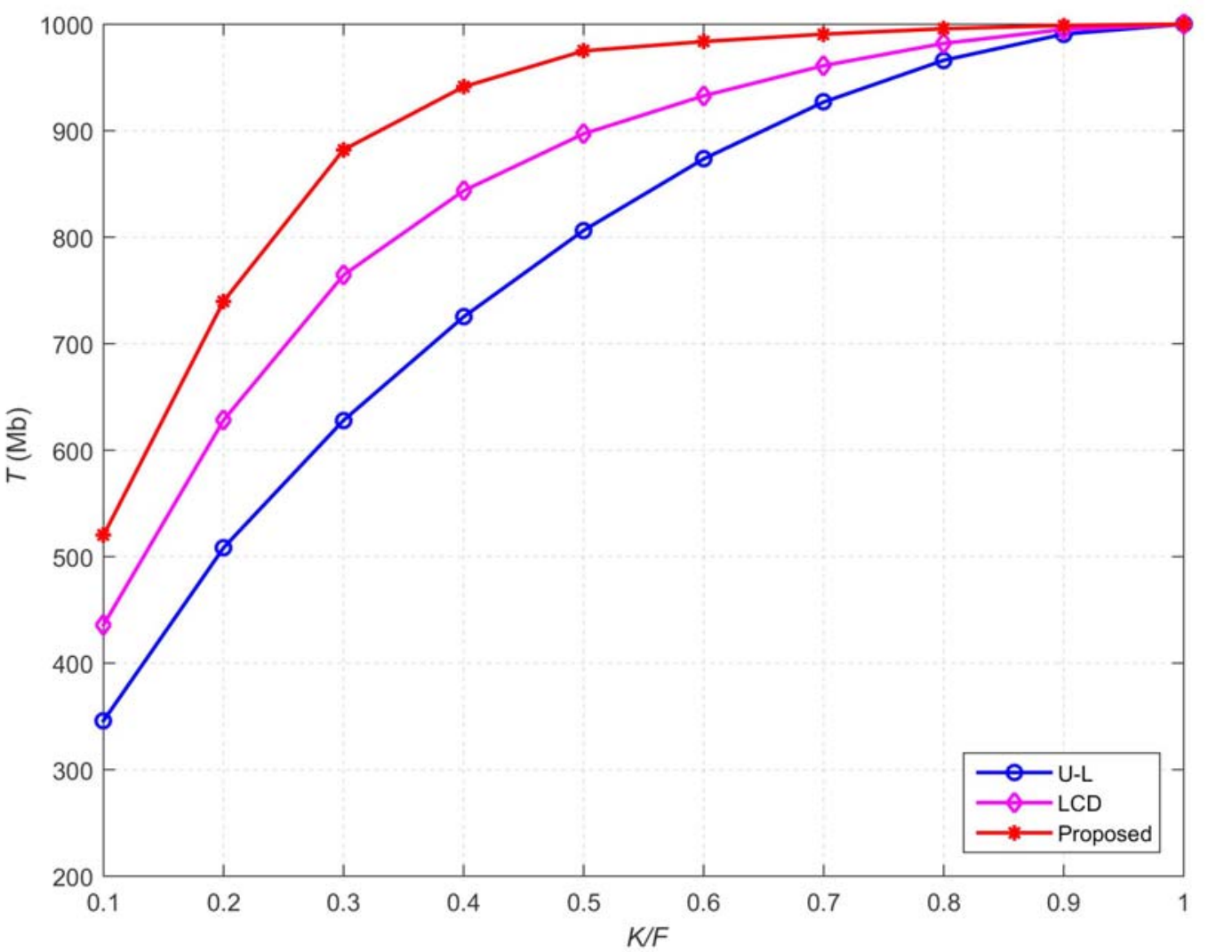}
\caption{The whole offloaded traffic versus the normalized storage size.}
\label{result1}
\end{figure}

\begin{figure}[!t]
\centering 
\includegraphics[width=0.48\textwidth]{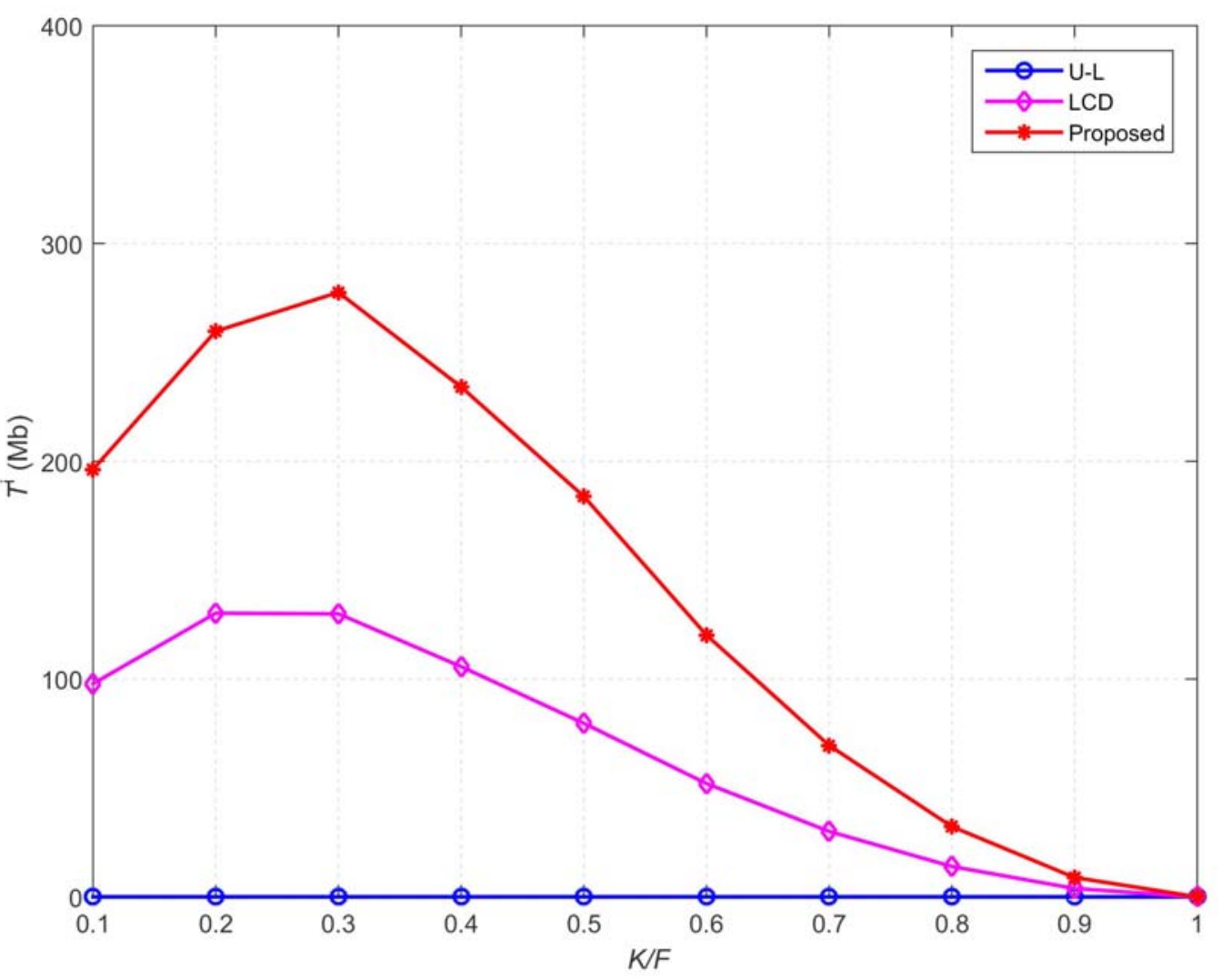}
\caption{The incremental  offloaded traffic versus the normalized storage size.}
\label{result2}
\end{figure}

\begin{figure}[!t]
\centering 
\includegraphics[width=0.48\textwidth]{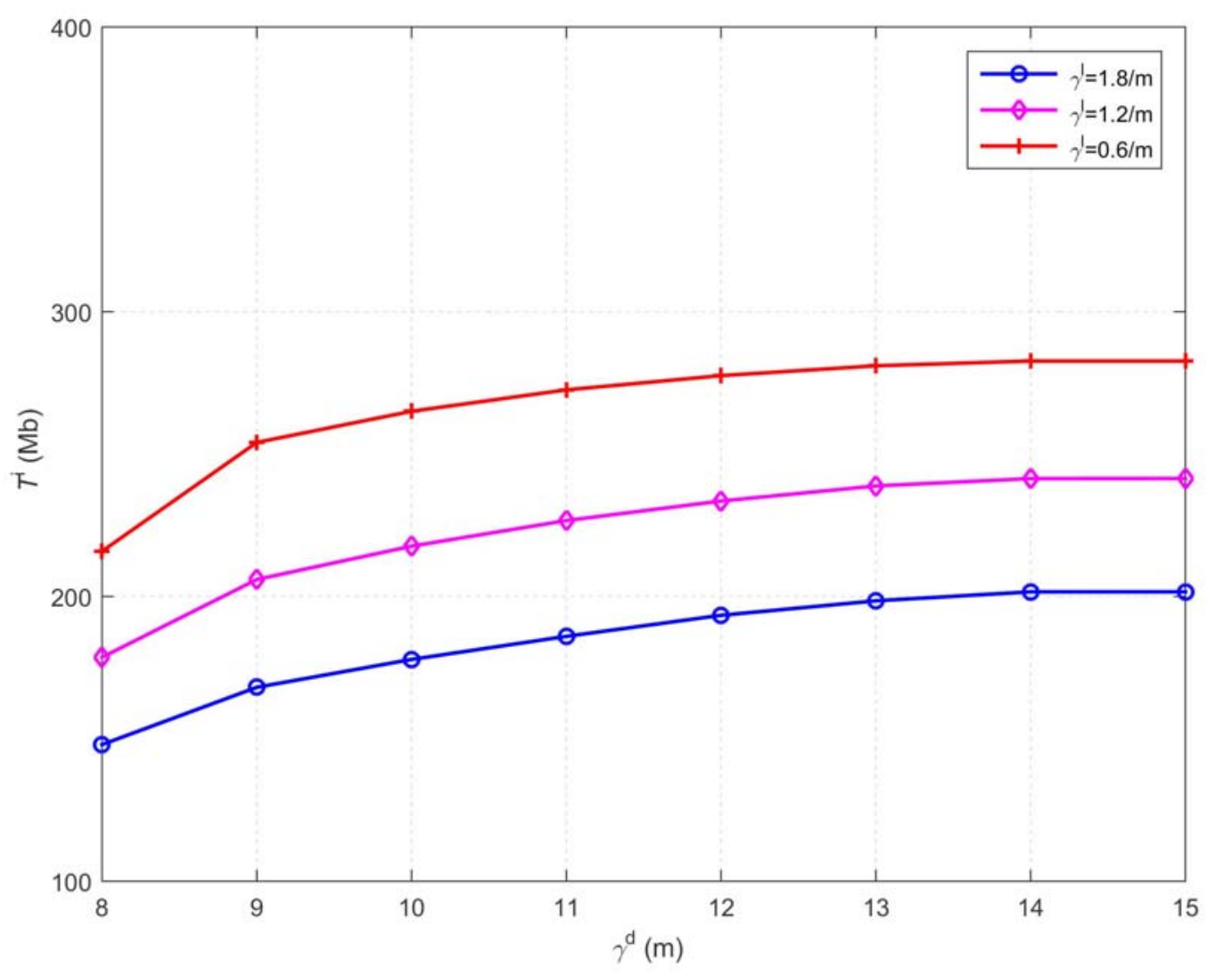}
\caption{The incremental  offloaded traffic versus the distance and load thresholds.}
\label{result3}
\end{figure}

In Fig. \ref{result1} and Fig. \ref{result2}, we show the whole offloaded traffics and the incremental offloaded traffics of our proposed approach and the two baseline schemes with different normalized storage size of each F-AP ($K/F$), respectively.
It can be observed that the whole offloaded traffics of all the three considered approaches increase with the normalized storage size.
It can also be observed the performance of the proposed approach is superior to  that of the baseline schemes.
The reason is that the proposed approach improves the way of clustering
with the consideration of local content popularity and reduces the repetitive and redundant storage  of files, correspondingly, more user requests can be responded locally  compared with the baseline schemes.

In Fig. \ref{result3}, we show the incremental offloaded traffic of our proposed approach  versus the distance threshold ${{\gamma ^\text{d}}}$ with different load threshold ${{\gamma ^{\text{l}}}}$.
It can be observed that the incremental offloaded traffic decreases with ${{\gamma ^{\text{l}}}}$ and increases with ${{\gamma ^\text{d}}}$.
The reason is that as ${{\gamma ^\text{d}}}$ becomes larger and ${{\gamma ^{\text{l}}}}$ becomes smaller, the constraints of the clustering optimization problem will be relaxed, and more F-APs can cooperative with each other.

\section{Conclusions}

In this paper, we have proposed a graph-based cooperative caching approach in F-RAN.
By constructing the relevant node graph and  weighted graph,
the objective cluster sets  have been obtained by searching for the max-weight independent subset of the vertex set of the weighted graph.
Both great computational complexity reduction and  remarkable offloading gain
have been achieved by using our proposed  approach.

\section*{Acknowledgments}

This work was supported in part
the National 863 Project (2015AA01A709),
the National Basic Research Program of China (973 Program 2012CB316004),
the Natural Science Foundation of China (61521061),
and the UK Engineering and Physical Sciences Research Council under Grant EP/K040685/2.

%

\bibliographystyle{IEEEtran}
\bibliography{manuscript-coopcaching-conf}

\end{document}